\documentclass{PoS}

\usepackage{multirow,lscape,amsmath,cite,amssymb,booktabs}

\title{Scale setting, sigma terms and the Feynman-Hellman theorem}

\ShortTitle{Scale setting, sigma terms and the Feynman-Hellman theorem}

\author{\speaker{P.E.~Shanahan}, A.W.~Thomas and R.D.~Young\\
        ARC Centre of Excellence in Particle Physics at the Terascale and CSSM, School of Chemistry and Physics, University of Adelaide,
  Adelaide SA 5005, Australia\\}

\abstract{The authors recently presented new values for the octet baryon sigma terms. These were extracted using the Feynman-Hellman theorem from a chiral perturbation theory fit to octet baryon mass data from the PACS-CS collaboration. Of particular interest is the precise determination of the strangeness sigma term $\sigma_s = 21 \pm 6$~MeV. In this work, we elaborate on the critical effect which the choice of scale setting has on this value. We discuss the prospect that the comparison of direct and `spectrum' determinations of the sigma terms from the lattice can provide insight not only into scale setting on the lattice, but into QCD itself.}

\FullConference{The 30th International Symposium on Lattice Field Theory\\
		 June 24 $-$ 29,  2012\\
		 Cairns, Australia}

\begin{document}

\section{The strange sigma term}

To quantify and investigate the nature of chiral symmetry breaking in QCD one would ideally like to measure objects such as the matrix element of the symmetry-breaking Hamiltonian density between nucleon states. While this is not possible directly, the sigma term can be obtained from on-shell scattering amplitudes.

The baryonic sigma terms are defined as scalar form factors:
for each quark flavor $q$ and baryon B,
\begin{equation}
\sigma_{Bq} = m_q \langle B | \overline{q} q | B \rangle.
\end{equation}
For the nucleon, the so-called $\pi N$ sigma commutator and the strange sigma commutator are defined as
\begin{align}
\sigma_{\pi N} & = m_l \langle N | \overline{u}u + \overline{d}d | N \rangle, \\
\sigma_s & = m_s \langle N | \overline{s}s | N \rangle,
\end{align}
where $m_l = (m_u+m_d)/2$.

Written in this form, it is clear that the sigma terms measure the strength of the various matrix elements $m\overline{q}q$ $(q =u, d, s)$ in the baryons. Hence, these terms quantify both the contribution of chiral symmetry breaking to the baryon masses, and the scalar density of quarks in the baryons.
As many dark matter candidates have interactions with hadronic matter which are determined by couplings to the light and strange quark sigma terms, these commutators are also of essential importance to the interpretation of experimental searches for a particle candidate of dark matter~\cite{Ellis:2008hf,Ellis2009,Giedt:2009mr,Bottino:1999ei,Hill:2011be,Underwood2012}.

Largely because of this importance to dark matter searches, recent years have seen a concerted effort towards precisely determining the nucleon sigma terms in particular. While the best value for $\sigma_{\pi N}$ has remained somewhat stable -- modern lattice calculations support a value $\sigma_{\pi N} \approx 45$~MeV as determined from $\pi N$ scattering through a dispersion relation analysis~\cite{Brown:1971,Gasser:1991} -- the best value for $\sigma_s$ has seen an enormous revision.  
Traditional indirect evaluations using $\sigma_{\pi N}$ and a best-estimate for the non-singlet contribution $\sigma_0= m_l\langle N|\overline{u}u + \overline{d}d - 2\overline{s}s|N \rangle$ have yielded a value for $\sigma_s$ as large as 300 MeV.
Since the strange sigma commutator may be interpreted as the contribution to the mass of the nucleon from the strange quark, a value as large as this would indeed be remarkable; it would suggest that almost a third of the nucleon mass arises from non-valence quarks. This appears incompatible with the widely used constituent quark models, for example. 
While early lattice simulations appeared to support large values of $\sigma_s$, advances in lattice QCD have revealed a strange sigma term of 20-50~MeV~\cite{Young:2009zb,Dinter:2012tt,Bali:2011ks,Dinter:2011zz,Horsley:2011wr,Durr:2011mp,Babich:2010at,Takeda:2010cw}, an order of magnitude smaller than was previously believed, and significantly more precise.
Recent work from QCDSF supports the suggestion that operator mixing is responsible for erroneously large values of $\sigma_s$ in the early lattice studies~\cite{Bali:2011ks}.


Shanahan~\textit{et al.}~\cite{Shanahan:2012wh} recently presented new values for the octet baryon sigma terms. These were extracted using the Feynman-Hellman theorem from a chiral perturbation theory fit to octet baryon mass data from the PACS-CS collaboration~\cite{Aoki:2008sm}. Of particular interest is the precise determination of the strangeness sigma term $\sigma_s = 21 \pm 6$~MeV. In this work, we elaborate on the critical effect which the choice of scale setting has on this value.

\subsection{Scale setting and the Feynman-Hellman theorem}

A common feature of lattice simulations is that all quantities are calculated in units of the unknown lattice spacing $a$, which must be determined by matching an observable to its physical value. This can be done in a variety of ways. Two common methods, often referred to as `mass independent' and `mass dependent' scale setting schemes, are of interest to us here:
\begin{enumerate}
\item The inverse bare coupling $\beta$ determines the lattice spacing $a$. That is, points at fixed $\beta$ are extrapolated to the physical point (usually linearly in the bare quark mass $am_{q_\textrm{sea}}$), and the value of some observable at that point is used to set the common scale $a$ for \textit{all} lattice ensembles at that common $\beta$.  
\item The lattice spacing varies with bare quark mass. That is, the lattice spacing $a$ is determined separately for each set of bare parameters $(\beta, am_{q_\textrm{sea}})$ by using a physical observable that is assumed to be independent of the quark masses (often the Sommer scale $r_0$). 
\end{enumerate}

We could think of these two choices as different ways of absorbing the observed quark mass dependence of the ratio $r_0/a$ at fixed $\beta$. Method 1 essentially assumes that this dependence may be attributed to the variation of $r_0$ with quark mass, while method 2 instead assumes that $a$ is changing. What is not often considered is that both $r_0$ and $a$ may have some dependence on the sea quark mass, which would lead to an intermediate scale in some sense. Such a `mixed' scale setting procedure would be non-trivial to implement.

Of course, in the continuum limit, and when the chiral extrapolation to physical quark masses has been performed, the results of each method of scale setting must agree for physical observables. When considering derived quantities which are expressed as derivatives with respect to quark mass, however, the choice of scale setting method becomes far more significant; these quantities, by the very definition of the derivative, depend on the scale away from the physical point and hence on the scale setting scheme chosen.
The Feynman-Hellman theorem expresses the baryon sigma terms precisely as such derivatives~\cite{Feynman:1939},
\begin{equation}
\label{eq:FH}
\sigma_{Bq} = m_q \frac{\partial M_B}{\partial m_q}.
\end{equation}

Clearly there are two primary methods of evaluating the baryon sigma terms on the lattice. The first is by direct calculation of the relevant matrix elements. This method does not suffer from any major scale setting ambiguities. The other is to use lattice data for the octet baryon mass spectrum and the Feynman-Hellman theorem to determine the sigma terms -- the `spectrum' method. The results of this method will depend on the choice of scale setting procedure, as discussed above. 

This effect may potentially be significant for the strange sigma term in particular.
One reason is the well-known `beta shift' effect which is observed when unquenching lattice simulations; at fixed $\beta$, the ratio $r_0/a$ increases when dynamical quarks are added~\cite{Bali:1993zj}, which can be interpreted as a sea quark dependence of either $r_0$, or the spacing $a$. As this shift can be significant, the choice of scale setting absorbs a possibly large effect, and hence will lead to non-negligible differences in the results of derivatives with respect to sea quark mass calculated with each of the two choices.
Additionally, the strange quark is considerably heavier than the light quarks, which, by Equation~(\ref{eq:FH}), serves to amplify the effect which a change in the slope of baryon mass with respect to strange quark mass has on the sigma term.

To illustrate the significance of the choice of scale setting method, we consider the recent determination of the octet baryon sigma terms presented by Shanahan~\textit{et al.}~\cite{Shanahan:2012wh}. This work used method 2 to set the scale.
If an identical analysis is repeated, using the same PACS-CS lattice data but with the scale set using method 1 instead, the results are significantly different. While the chiral extrapolations of the octet baryon masses are compatible at the physical point, as expected, and the qualities of the two fits are similar ($\chi^2$/d.o.f 0.41 and 0.67 for methods 2 and 1), the value of the strangeness sigma term changes significantly. In particular, we note that the value for $\sigma_s$ in the nucleon changes from 21$\pm$6~MeV to 59$\pm$7~MeV as the scale setting scheme is changed from method 2 to method 1. Despite the small statistical and model-dependent uncertainties of the calculation, there is clearly a significant dependence on the scale setting method. Results for the light quark sigma terms and for other octet baryons are given in tables~\ref{table:sigmasOrig} and~\ref{table:sigmasOther}. For each baryon, there is a shift of about 35~MeV in $\sigma_{Bs}$. Clearly, a deeper understanding of scale setting on the lattice is of essential importance for the interpretation of spectral determinations of the sigma terms. 

\begin{center}
\begin{table}[tbh]
\begin{center}
\begin{tabular}{c c c c c}
$B$ & Mass (GeV) & Experimental & $\sigma_{Bl}$~(GeV) & $\sigma_{Bs}$~(GeV) \\ 
\midrule
$N$ & 0.959(24)(9) & 0.939 & 0.045(5)(5)  & 0.021(6)(0) \\
$\Lambda$ & 1.129(15)(6) & 1.116 & 0.029(3)(2) & 0.159(7)(1) \\
$\Sigma$ & 1.188(11)(6) & 1.193 & 0.024(2)(2) & 0.204(8)(1) \\
$\Xi$ & 1.325(6)(2) & 1.318 & 0.0117(9)(5) & 0.316(9)(2) \\
\end{tabular}
\end{center}
\caption{Extracted masses and sigma terms for the physical
  baryons, with the lattice scale set using method 2. The first uncertainty quoted is statistical, while the
  second results from the variation of various chiral parameters and
  the form of the UV regulator as described in~\cite{Shanahan:2012wh}, from which these results are taken. The
  experimental masses are shown for comparison.}
\label{table:sigmasOrig}
\end{table}
\end{center}

\begin{center}
\begin{table}[tbh]
\begin{center}
\begin{tabular}{c c c c c}
$B$ & Mass (GeV) & Experimental & $\sigma_{Bl}$~(GeV) & $\sigma_{Bs}$~(GeV) \\ 
\midrule
$N$ & 0.944(22)(9) & 0.939 & 0.051(3)(6)  & 0.059(6)(1) \\
$\Lambda$ & 1.119(15)(6) & 1.116 & 0.034(2)(2) & 0.193(8)(1) \\
$\Sigma$ & 1.182(11)(6) & 1.193 & 0.028(2)(2) & 0.241(9)(2) \\
$\Xi$ & 1.323(7)(2) & 1.318 & 0.0151(9)(4) & 0.352(11)(1) \\
\end{tabular}
\end{center}
\caption{Extracted masses and sigma terms for the physical baryons, with the scale set using method 1. The uncertainties are as in 
table~\protect\ref{table:sigmasOrig}.  }
\label{table:sigmasOther}
\end{table}
\end{center}

The effect of the choice of scale setting on such spectral determinations of the sigma terms has been discussed in the literature. Both \cite{De:2008xt} and \cite{Durr:2010ni} argue for method 1, in the former case based on observations of scaling violation, while others, for example~\cite{Foster:1998vw,Michael:2001bv} instead favour method 2. There is no clear consensus as to the most appropriate way to set the scale.

In the work referenced here~\cite{Shanahan:2012wh}, Shanahan~\textit{et al.} chose method 2 to set the scale for their primary analysis, using the argument that the Sommer scale $r_0$, based on the force between static quarks at some distance, should be independent of small changes in the (small) bare quark masses.
This approach is strongly supported by the fact that the results of this calculation agree with~\textit{direct} lattice calculations at unphysical quark masses. For example, as discussed in~\cite{Shanahan:2012wh}, the QCDSF Collaboration~\cite{Bali:2011ks} recently presented a value $\sigma_{s_{\textrm{QCDSF}}}=12^{+23}_{-16}$~MeV at $(m_\pi,m_K)=(281,547)$~MeV which agrees well with the results of the calculation of~\cite{Shanahan:2012wh} with the scale set using method 2, namely $\sigma_{s_2}=16(5)(1)$~MeV. In contrast, the direct result is not consistent with the value $\sigma_{s_1}=56(6)(1)$~MeV obtained when the scale is set using method 1. Similarly, other recent direct lattice calculations favour smaller values of $\sigma_s$~\cite{Oksuzian:2012rza,Dinter:2012tt}. While~\cite{Babich:2010at} present a larger $\sigma_s$, naively more compatible with scale setting method 1, the authors point out that the effects of operator mixing likely serves to increase their result, again suggesting a true value more compatible with method 2. We point out that while most recent direct results present small values for $\sigma_s$, not all calculations do support this trend~\cite{Freeman:2012ry,Engelhardt:2012gd}. 

Despite this debate, we suggest that method 2 of setting the lattice scale yields results which are more consistent with direct calculations.
Of course, we cannot rule out the possibility of `mixed scale setting' as discussed earlier, and the precision of the result of $\sigma_s =21 \pm 6$~MeV relies heavily on the assumption that $r_0$ does not change with quark mass.
We emphasize that as more precise direct lattice calculations for the strangeness sigma commutator appear, we will not only more precisely determine the value of this term as needed for dark matter calculations, but we stand to gain significant insight into the problem of scale setting on the lattice and indeed into QCD itself.

\section{Up, Down and Strange Sigma Terms}

As discussed in previous sections, Shanahan~\textit{et al.}~\cite{Shanahan:2012wh} recently presented new values for the light and strange quark octet baryon sigma terms. This work applied the Feynman-Hellmann relation to the results of a chiral extrapolation of PACS-CS Collaboration~\cite{Aoki:2008sm} lattice data for octet baryon masses. Arguably the most significant result presented was a small value for the strangeness sigma term, namely $\sigma_s = 21 \pm 6$~MeV at the physical point.
In what follows, we summarize additional results of this work which were not presented in the original paper. In particular, using the approach presented in~\cite{Shanahan:2012wa}, individual up, down and strange sigma terms as relevant to supersymmetric dark matter searches~\cite{Ellis:2008hf,Ellis:2000jd,Ellis:2000ds} may be evaluated.

In reference~\cite{Shanahan:2012wa}, the authors used the same chiral perturbation theory fits to octet baryon mass data as was used in their evaluation of the sigma terms~\cite{Shanahan:2012wh}. By extending the chiral extrapolation to the case of broken isospin symmetry, i.e., $m_u \ne m_d$, they determined the mass splittings between members of baryonic isospin multiplets based on their fit to isospin-averaged lattice data.
We use precisely the same method to extract the individual, isospin-broken, sigma terms $f_{Tu} \ne f_{Td}$. Just as was done in~\cite{Shanahan:2012wa}, we take as input the physical up-down quark mass ratio~\cite{Leutwyler1996},
\begin{equation}
R:=\frac{m_u}{m_d} = 0.553 \pm 0.043,
\label{eq:R}
\end{equation}
determined by a fit to meson decay rates. We note that this value is compatible with more recent estimates of the ratio from $2+1$ and 3 flavor QCD and QED~\cite{Aubin2004,Blum2010}. All calculations include in quadrature the uncertainty due to the stated range for $R$.

Following the notation of Ellis~\textit{et al.}~\cite{Ellis:2008hf,Ellis:2000jd,Ellis:2000ds}, we define individual quark ($q$) sigma terms in the proton as
\begin{equation}
m_p f_{Tq}^{(p)} \equiv \langle p | m_q \overline{q}q | p \rangle \equiv m_q B_q,
\end{equation}
and also define the related quantity
\begin{equation}
z \equiv \frac{B_u-B_s}{B_d-B_s}.
\end{equation}
These terms are related to the previously defined $\sigma_s$ and $\sigma_{\pi N}$ by
\begin{align}
\sigma_{\pi N} & = \frac{1}{4}(m_u+m_d)(B_u^{(p)}+B_d^{(p)}+B_u^{(n)}+B_d^{(n)}) \\
\sigma_s & = \frac{1}{2}m_s(B_s^{(p)}+B_s^{(n)}).
\end{align}

The results of our analysis are
\begin{align}
f_{Tu}^{(p)}&=0.019(3), & f_{Td}^{(p)}& =0.027(4), & f_{Ts}^{(p)}=0.023(7), \\
f_{Tu}^{(n)}&=0.015(2), & f_{Td}^{(n)}& =0.033(5), & f_{Ts}^{(n)}=0.022(6).
\end{align}
We also find
\begin{align}
z &= 1.27(3), &\frac{B_d}{B_u} &= 0.79(2), & \frac{m_l}{M_N}\langle N | \overline{u}u - \overline{d}d | N \rangle =0.0049(2).
\end{align}
The quoted errors include correlated uncertainties between all fit parameters, and allow for some variation of phenomenologically set quantities including the up-down quark mass ratio $R$. For details see~\cite{Shanahan:2012wh,Shanahan:2012wa}.

These results may be compared with those in~\cite{Ellis:2000ds} (we note that the same $R$ value was used in that analysis). While our up and down sigma terms are compatible, within uncertainties, with these results, we point out that the strange sigma terms $f_{Ts}$ and the parameter $z$ resulting from our work are significantly smaller than the previous values.

Of course, the discussion of the previous section as regards $\sigma_s$ applies equally to $f_{Ts}$: setting the lattice scale with fixed coupling $\beta$ corresponding to a fixed lattice spacing yields far larger values for the strange sigma terms, namely $f_{Ts}^{(p)} = 0.063(7)$ and $f_{Ts}^{(n)}=0.062(7)$, but a compatible value $z=1.25(2)$.

\acknowledgments 

This work was supported by the University of Adelaide and the Australian Research Council through the ARC Centre of Excellence for Particle Physics at the Terascale and grants FL0992247 (AWT) and DP110101265 (RDY).

\end{document}